# Engineering RAG Systems for Real-World Applications: Design, Development, and Evaluation

Md. Toufique Hasan(✉), Muhammad Waseem, Kai-Kristian Kemell, Ayman Asad Khan, Mika Saari, and Pekka Abrahamsson

Faculty of Information Technology and Communication Sciences, Tampere University, Tampere, Finland
{mdtoufique.hasan,muhammad.waseem,kai-kristian.kemell,ayman.khan, mika.saari,pekka.abrahamsson}@tuni.fi

**Abstract.** Retrieval-Augmented Generation (RAG) systems are emerging as a key approach for grounding Large Language Models (LLMs) in external knowledge, addressing limitations in factual accuracy and contextual relevance. However, there is a lack of empirical studies that report on the development of RAG-based implementations grounded in real-world use cases, evaluated through general user involvement, and accompanied by systematic documentation of lessons learned. This paper presents five domain-specific RAG applications developed for real-world scenarios across governance, cybersecurity, agriculture, industrial research, and medical diagnostics. Each system incorporates multilingual OCR, semantic retrieval via vector embeddings, and domain-adapted LLMs, deployed through local servers or cloud APIs to meet distinct user needs. A web-based evaluation involving a total of 100 participants assessed the systems across six dimensions: (i) Ease of Use, (ii) Relevance, (iii) Transparency, (iv) Responsiveness, (v) Accuracy, and (vi) Likelihood of Recommendation. Based on user feedback and our development experience, we documented twelve key lessons learned, highlighting technical, operational, and ethical challenges affecting the reliability and usability of RAG systems in practice.

## 1 Introduction

Retrieval-Augmented Generation (RAG) is an approach that integrates external knowledge retrieval into language model outputs to improve accuracy and relevance. RAG enhances Large Language Models (LLMs) by retrieving relevant external knowledge, thereby strengthening the performance of Generative AI (GenAI) applications. While GenAI has seen use in software engineering [19], RAG extends its value to broader domains by combining parametric and non-parametric memory, effectively addressing the limitations of static knowledge bases [27]. Recent advances in retrieval-augmented LLMs enable real-time information retrieval, reducing hallucinations and improving response reliability [7].

The foundational work by Lewis et al. [14] established RAG as a standard for tasks like question answering and knowledge retrieval. Early frameworks such as REALM and RAG demonstrated the benefits of combining dense retrieval with language generation for open-domain tasks [10]. However, most later research has focused on improving retrieval architectures and reducing hallucinations, often evaluated only on clean, English-centric benchmarks [6]. There remains limited exploration of domain-specific, multilingual, or real-world deployments, which this paper addresses through the development and evaluation of RAG systems focused on retrieval quality and system design.

Accurate information access is important in domains like governance, cybersecurity, agriculture, industrial research, and healthcare. As industries adopt AI for complex tasks, traditional search methods often fall short, especially with multilingual, up-to-date, and contextually relevant knowledge. To address this, we developed RAG systems in collaboration with five organizations: the City of Kankaanpää[1], Disarm[2], AgriHubi[3], FEMMa[4], and a clinical diagnostics group[5]. Each collaboration guided the system design to meet distinct operational and information access challenges.

This study investigates how RAG systems can be engineered and evaluated in real-world contexts, focusing on system design, domain-specific applications, user evaluation, and lessons for future practice. Guided by these aims, this paper addresses the following Research Questions (RQs):

- **RQ1:** How can RAG systems be designed and developed to address real-world system needs across diverse application domains, and what are the lessons learned from engineering RAG systems for real-world applications?
- **RQ2:** How do users evaluate domain-specific RAG systems in terms of ease of use, relevance, transparency, responsiveness, and accuracy in real-world applications?

To address these research questions, we developed five domain-specific RAG systems and evaluated them through a web-based user study with 100 participants. The evaluation focused on usability, retrieval relevance, transparency, and other user-centred factors. Full methodological details are provided in Sect. 3.

The contributions of this paper are as follows: end-to-end development and deployment of RAG systems for multilingual, domain-specific applications; user-centred evaluation demonstrating real-world performance across usability and accuracy metrics; practical engineering insights to guide the design of reliable and maintainable RAG pipelines; and system-level considerations for integrating

[1] https://www.kankaanpaa.fi/.
[2] https://www.disarm.foundation/framework.
[3] https://maaseutuverkosto.fi/en/agrihubi/.
[4] https://www.tuni.fi/en/research/future-electrified-mobile-machines-femma.
[5] https://tampere.neurocenterfinland.fi/.

RAG into real-world AI-based software, contributing to software engineering practice.

**Paper Structure:** Section 2 reviews related work. Section 3 outlines the study design. Section 4 details the system implementation. Section 5 presents the evaluation. Section 6 highlights key lessons learned. Section 7 discusses limitations, and Sect. 8 offers conclusions and future directions.

## 2 Related Work

RAG improves the factual accuracy and contextual relevance of LLMs by incorporating real-time external information, making it especially valuable for complex tasks such as question answering, legal reasoning, and summarization [20]. Recent work has demonstrated RAG's utility in taxonomy-driven dataset design [16], token-efficient document handling [20], and multimodal applications that combine text and images via Vision-Language Models (VLMs) such as VISRAG [29]. Despite these advances, OCR noise remains a limiting factor in retrieval fidelity [32]. Ongoing research addresses this by refining dataset construction [16], tackling architectural scalability [5], improving query-document alignment through prompt engineering [34], and applying speculative retrieval to boost performance in multimodal settings [33].

RAG has been applied in software engineering for code understanding and developer tasks, e.g., StackRAG [1] enhances developer assistance using Stack Overflow, and CodeQA [2] applies LLM agents with retrieval augmentation. Ask-EDA [22] reduces hallucinations in Electronic Design Automation via hybrid retrieval. In industry, Khan et al. [12] address PDF-focused retrieval, while Xiaohua et al. [26] propose re-ranking and repacking for pipeline optimization. In healthcare, MEDGPT [24] extracts structured diagnostic insights, Path-RAG [18] improves pathology image retrieval, Alam et al. [3] introduce multi-agent retrieval for radiology reports, and Guo et al. [9] present LightRAG for graph-based precision retrieval.

RAG continues to expand into domains like energy and finance. Gamage et al. [8] propose a multi-agent chatbot for decision support in net-zero energy systems, while HybridRAG [21] combines knowledge graphs with vector search to enhance financial document analysis. AU-RAG by Jang and Li [11] dynamically selects retrieval sources using metadata, improving adaptability across sectors. To address retrieval noise, Zeng et al. [31] integrate contrastive learning and PCA for better knowledge filtering. Barnett et al. [4] identify core RAG weaknesses, including ranking errors and incomplete integration, underscoring the ongoing need for more reliable retrieval strategies.

The rise of autonomous AI agents has further improved RAG by enabling self-directed reasoning, adaptive retrieval, and memory persistence. Wang et al. [25] survey LLM-driven agent architectures, while Liu et al. [17] benchmark multi-turn reasoning through Agent-bench. Agent-tuning by Zeng et al. [30]

enhances instruction tuning for retrieval-based decisions. Singh et al. [23] categorize Agentic RAG into single-agent, multi-agent, and graph-based designs, highlighting dynamic tool use. On the retrieval side, Yan et al. [28] introduce CRAG to reduce hallucinations using confidence-based filtering, and Li et al. [15] improve precision through contrastive in-context learning and focus-mode filtering, strengthening RAG's reliability in complex scenarios.

**Conclusive Summary:** While RAG continues to advance, challenges in retrieval accuracy, reliability, and scalability persist [13]. Despite progress in hybrid strategies [21], autonomous agents [23], and correction techniques [15,28,31], domain-specific evaluation remains limited. This paper evaluates five RAG systems and offers practical development and deployment insights.

## 3 Study Design

Figure 1 provides an overview of the methodological steps, detailing the phased implementation of five domain-specific RAG systems, a user-centred evaluation with 100 participants, and the synthesis of lessons learned across technical, operational, and ethical dimensions.

### 3.1 Implementing RAG Systems

This section describes how we designed, and built the RAG systems featured in this study. It explains the overall system design, how we selected the case study domains, the unique challenges each domain presented, and the setup used for evaluation.

**Domain Selection.** We selected the application domains to test RAG systems in real-world, knowledge-heavy environments where accurate information retrieval, contextual understanding, and timely decision-making are important. These domains were chosen because they involve different information and require careful decision-making, providing a solid basis to evaluate how well RAG systems can adapt and perform in different settings.

In this study, we apply RAG across five domains: municipal governance, cybersecurity, agriculture, industrial research, and medical diagnostics, to explore how RAG-based retrieval can address diverse domain-specific information needs and support real-world decision-making processes.

**System Design.** The design of the RAG systems in this study follows a two-phase approach:

– *Retrieval Phase:* User queries are embedded using pre-trained models (e.g., `text-embedding-ada-002`) and matched with relevant text chunks via similarity search in vector databases.

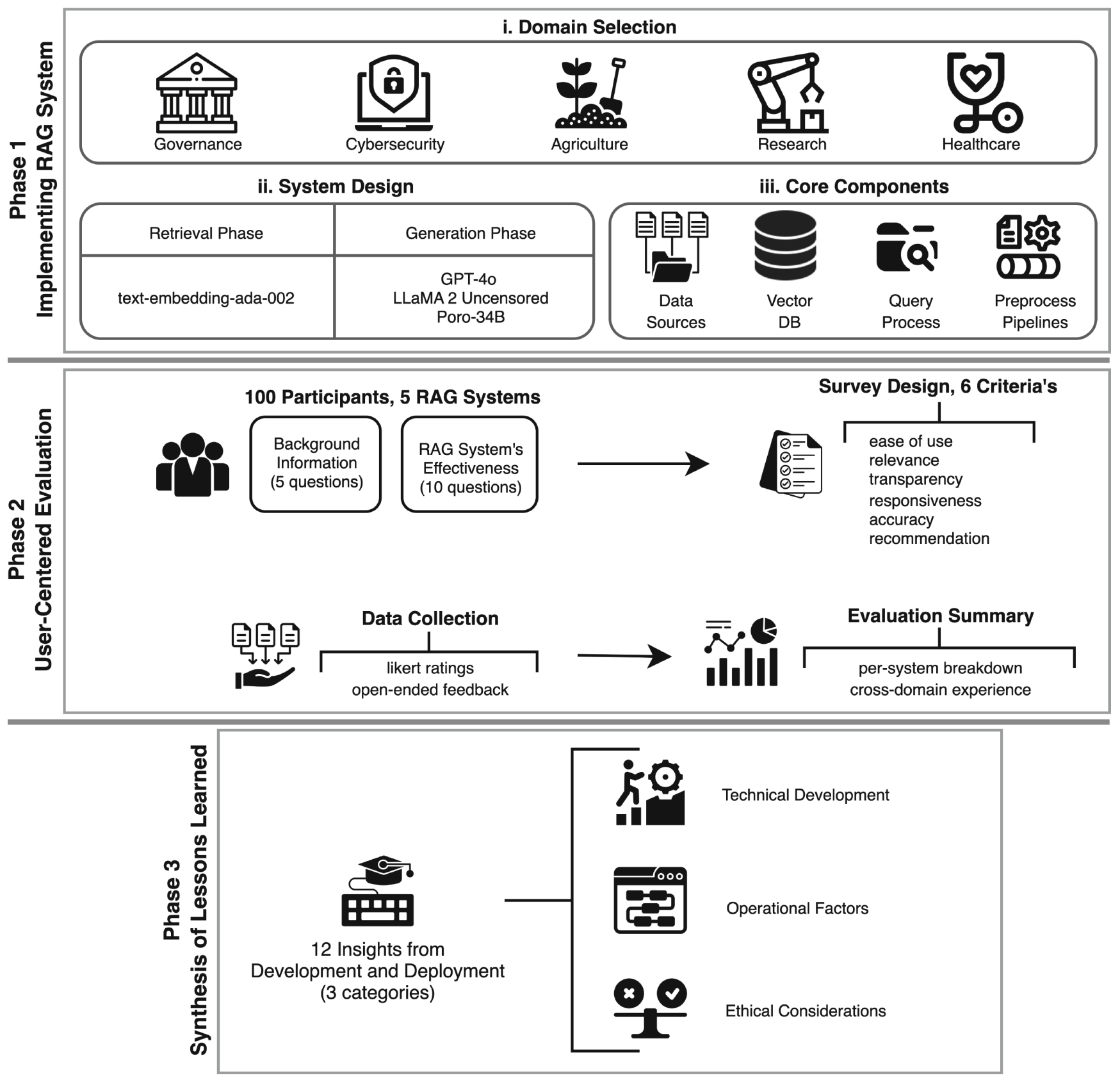

**Fig. 1.** Overview of the research methodology

– *Generation Phase:* The retrieved text chunks are concatenated with the original user query and passed into a large language model (LLM), such as `GPT-4o`, `LLaMA 2 Uncensored`, or `Poro-34B`, to synthesize contextually relevant responses.

This approach improves factual accuracy, minimizes hallucinations, and delivers insights that are well aligned with domain-specific needs.

**Core Components.** Each RAG-based system comprises multiple core components:

– *Data Sources:* Knowledge bases include structured and unstructured documents, such as websites, municipal records, cybersecurity reports, agricultural research papers, engineering documents, and clinical guidelines.
– *Vector Database:* The retrieved knowledge is stored as vector embeddings in `FAISS`, `Pinecone`, or `OpenAI's Vector Store`, depending on the system's latency and scalability requirements.

– *Query Processing:* User queries undergo tokenization, embedding generation, and similarity search before being passed to an LLM for the response.
– *Preprocessing Pipelines:* Systems rely on `PyMuPDF` and `Tesseract OCR` to extract text from PDFs and scanned documents, ensuring the inclusion of both text-based and image-based content. Additionally, for web scraping, the pipeline utilizes `BeautifulSoup`, `Scrapy`, and `Selenium` to extract, clean, and structure data from dynamic and static web pages.

These components enable efficient retrieval and context-aware responses in domain-specific RAG systems.

### 3.2 System Evaluation Method

To understand how the RAG-based systems performed in real usage scenarios, we conducted a structured web-based user study with 100 participants. Each participant was given access to live demo environments and interacted with one or more of the five systems using realistic, domain-specific tasks.

After using the systems, participants completed a standardized survey covering six criteria: Ease of Use, Relevance of Info, Transparency, System Responsiveness, Accuracy of Answers, and Recommendation. The survey included both Likert-scale questions and open-ended feedback. This approach provided both quantitative ratings and qualitative insights into system performance. We reviewed the open-ended feedback to identify common themes in participants' experiences. We also referred to development notes taken throughout the project. These helped us recognize recurring issues and informed the lessons described in Sect. 6.

## 4 Systems Implementation

This section outlines the implementation of five RAG-based systems for real-world deployment across diverse domains. Figure 2 presents the system architecture, showing user interaction, domain-specific data processing, vector storage, retrieval, and response generation with LLMs.

1. **Kankaanpää City AI**: This system enhances transparency of government records. It processes over 1,000 PDFs from 2023–2024, indexing them in `FAISS` for accurate retrieval of policy documents. The system uses the embedding model `text-embedding-ada-002` to convert documents into vector representations, and `gpt-4o-mini` as the LLM to generate context-aware responses. This setup allows users to search and access municipal decisions, infrastructure projects, and public policies with ease.
2. **Disarm RAG**: It is designed to deliver real-time insights into cyber threats, and forensic investigations. The system uses `LLaMA 2-uncensored` via Ollama to enable open access to cybersecurity knowledge, and is hosted on a secure server at CSC[6] (Finnish IT Center for Science) to ensure full data privacy. The system integrates red team techniques (e.g., phishing, deep-fake disinformation, privilege escalation) and blue team strategies (e.g., bot detection, misinformation control, network forensics), grounded in the `Disarm Framework`. It supports queries such as "How would you create a deep-fake to discredit a public figure?" and "What are the latest techniques for bypassing multi-factor authentication (MFA)?", as well as defensive questions like "How would you detect a disinformation campaign early on?" and "What are effective countermeasures against deep-fake-based phishing attacks?".

[6] https://research.csc.fi/cloud-computing/.

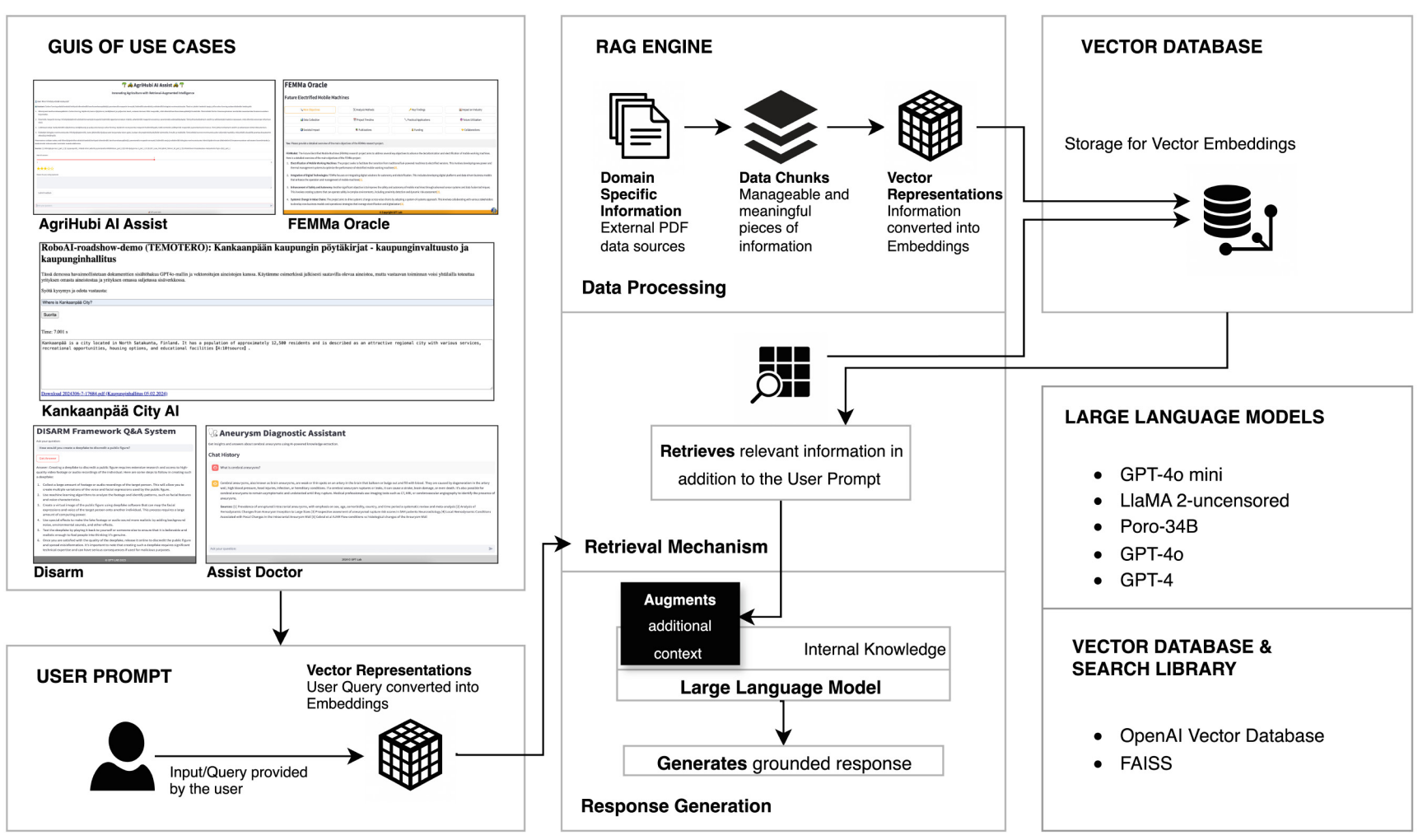

**Fig. 2.** System architecture of five RAG-based systems, showing data processing, vector storage, retrieval, and LLM-based response generation.

3. **AgriHubi AI Assist**: AgriHubi bridges agricultural policy and practice by processing 200+ Finnish-language PDFs using multilingual OCR and embedding the content into a `FAISS` vector database. It leverages the Finnish-optimized `Poro-34B` language model to deliver contextually relevant responses on topics like sustainable farming and soil conservation. The system features a `Streamlit` chat interface, logs interactions via `SQLite`, and includes a feedback mechanism for continuous improvement, making agricultural knowledge more accessible to farmers and researchers.
4. **FEMMa Oracle**: This system optimizes knowledge retrieval for engineering research, particularly in electrified mobile machinery. It processes around 28 PDFs regarding electrified mobile machinery. It integrates `GPT-4o` and `text-embedding-3-large` with `OpenAI's Vector Store` to enable rapid retrieval of structured engineering research documents. The system ensures that researchers can efficiently access validated technical documentation and structured project information, improving efficiency in engineering-related knowledge retrieval.

5. **Assist Doctor**: It is an aneurysm diagnostic RAG based application, developed at `Tampere University` for use by neurologists, radiologists, and vascular surgeons. It retrieves insights from peer-reviewed literature and clinical data using an embedding-based search pipeline and delivers context-aware responses via OpenAI's `GPT-4`. With a `Streamlit` interface, it enables clinicians to access diagnostic criteria, risk stratification models, and treatment comparisons, supporting informed decisions in aneurysm care.

All developed systems comply with *GDPR* and display source references, except *Disarm RAG*, where citations are hidden for security reasons.

## 5 Systems Evaluation

Understanding the real-world effectiveness of RAG-based systems requires moving beyond technical benchmarks to incorporate user-centred evaluation. We conducted a structured user study across five domain-specific deployments, capturing both system performance metrics and user perceptions of trust, relevance, and usability. This practical feedback offers a grounded view of system behaviour in real settings and highlights opportunities for targeted improvements.

### 5.1 Participant Demographics and RAG Orientation

To contextualize the system evaluation, we collected detailed background information from the 100 participants involved in the study. Figure 3 illustrates five key dimensions of participant orientation relevant to domain-specific RAG systems: professional role, AI vs. manual search preference, familiarity with RAG, prior usage experience, and comfort with AI-generated outputs.

1. **Role Distribution:** Participants represented five distinct professional categories aligned with our target application domains. Researchers comprised the largest segment (44%), followed by students (20%), domain experts (17%), AI/ML practitioners (16%), and others (3%). This composition reflects a balanced blend of technical stakeholders and domain users, ensuring that the evaluation captures both system-level performance and practical applicability across real-world contexts.
2. **AI-Generated vs. Manual Document Search:** Participants exhibited a task-sensitive perspective on AI assistance. While a substantial majority (83%) preferred AI-generated responses depending on the nature of the task, only (9%) expressed a consistent preference for AI over manual methods. Conversely, (8%) favoured manual search regardless of context. These findings suggest that trust in RAG systems is not absolute but contingent–underscoring the importance of response relevance, transparency, and alignment with user intent.
3. **Familiarity with AI-Based RAG:** Participants demonstrated a strong familiarity with RAG technologies in general, with (75%) identifying as either moderately (41%) or very familiar (34%) with AI-based RAG systems. However, since many participants were not domain experts in the specific fields covered by the systems (e.g., healthcare, cybersecurity), their feedback primarily reflects their interaction experience with RAG rather than deep subject-matter validation.

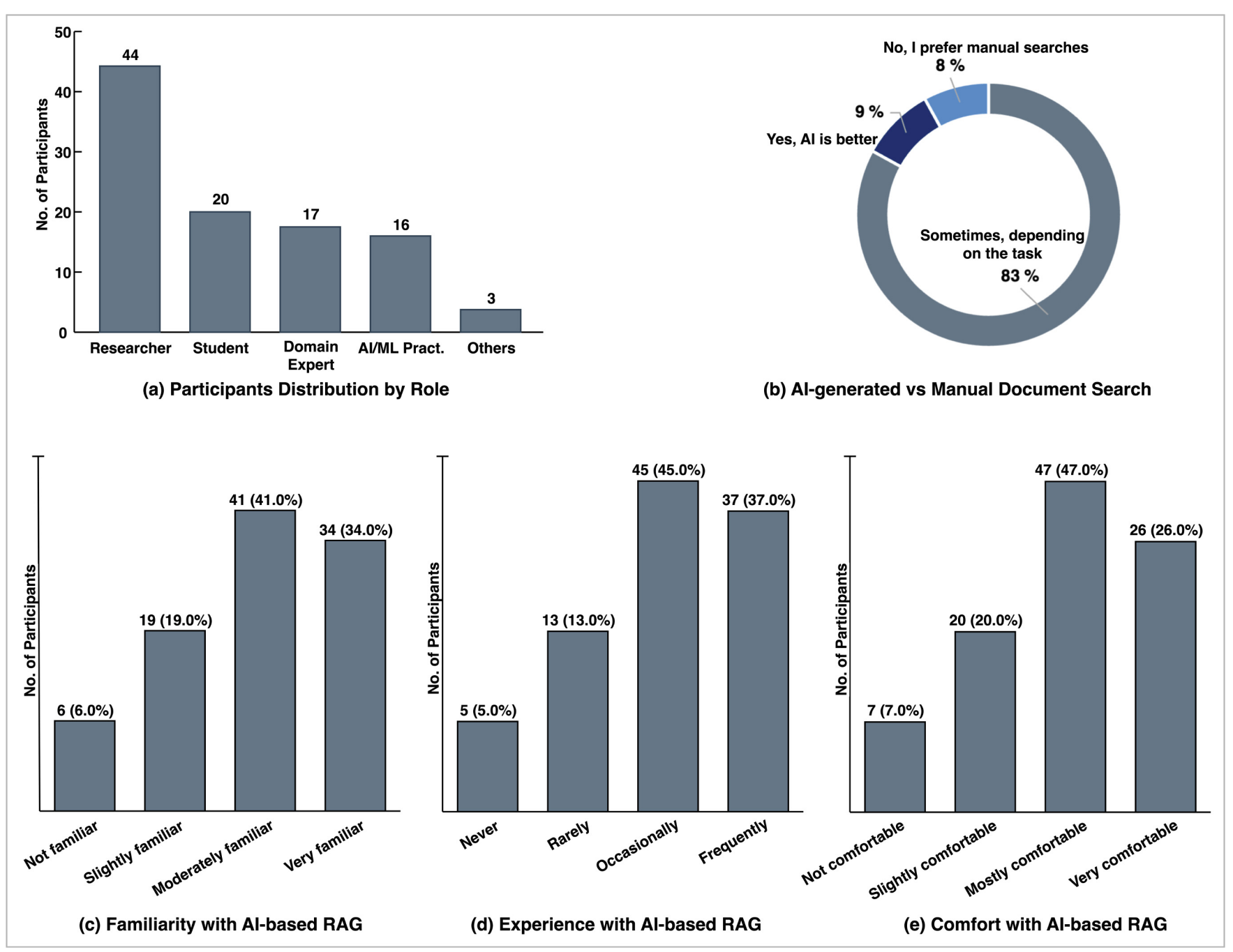

**Fig. 3.** Participant profiles and interaction with the RAG systems.

4. **Experience with AI-Based RAG:** Participant engagement with RAG systems was notably high. A majority (82%) reported using such systems either occasionally (45%) or frequently (37%), while only (5%) indicated no prior experience. This distribution reinforces the reliability of the feedback collected, as most evaluations were informed by direct, hands-on interaction rather than hypothetical exposure.
5. **Comfort with AI-Generated Responses:** Overall, participants expressed high confidence in AI-generated outputs. Nearly three-quarters (73%) reported feeling either mostly (47%) or very comfortable (26%) relying on such responses. Only a small minority (7%) expressed discomfort, indicating a strong baseline of user trust and an encouraging signal for broader adoption of generative AI in domain-specific tasks.

### 5.2 Survey Instrument and Case-Wise Findings

Figure 4 presents the aggregated user ratings across six evaluation criteria for all five RAG systems, offering a comparative perspective on system performance.

To capture both measurable and descriptive insights, we employed a survey combining Likert-scale questions (1–5 scale) with open-ended prompts for qualitative feedback. The evaluation focused on the following six core dimensions:

– *Ease of Use:* How easy was it to use the system?
– *Relevance of Information:* Did the system retrieve relevant and useful information for your queries?
– *Transparency:* Did the system show where the information came from?
– *System Responsiveness:* How would you rate the system's responsiveness in retrieving answers?
– *Accuracy of Answers:* Based on your knowledge, how accurate were the AI-generated answers provided by the system?
– *Recommendation:* Would you recommend this tool to colleagues in your field?

All five RAG systems were evaluated using the same six criteria by a total of 100 participants. The summaries below reflect how each system performed, highlighting key strengths and areas for improvement.

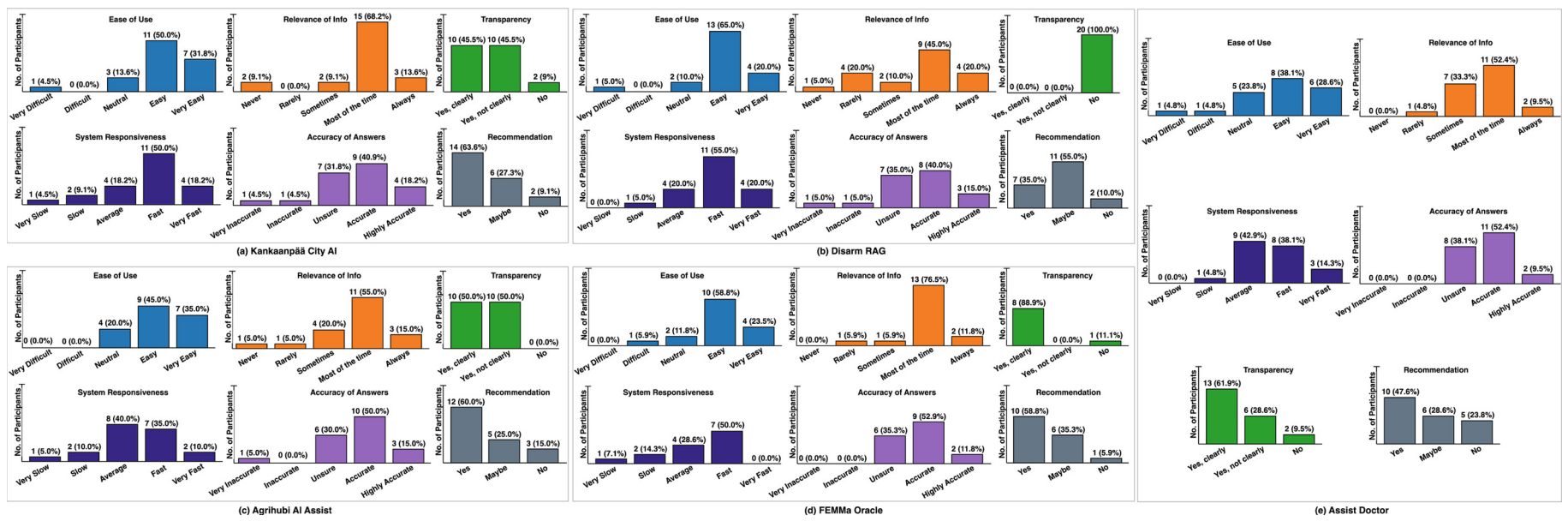

**Fig. 4.** User ratings of five RAG systems across six evaluation criteria.

1. **Kankaanpää City AI (22 participants):** The system performed well in Ease of Use, with (81.8%) rating it as "easy" or "very easy." Relevance of Info around (82%) and Accuracy of Answers around (91%) were also strong. Transparency was mixed, (45.5%) found it clear, while another (45.5%) found it unclear. (63.6%) said they would recommend the system, suggesting it may be useful in public governance contexts.
2. **Disarm RAG (20 participants):** Participants reported positive ratings for Ease of Use (65%) and System Responsiveness (75%), despite the complexity of the cybersecurity domain. Relevance of Info and Accuracy of Answers received moderate ratings, while Transparency was low due to intentionally hidden sources. Nevertheless, (55%) of participants indicated they would recommend the system.

3. **AgriHubi AI Assist (20 participants):** Tailored for Finnish-language agricultural content, the system received strong ratings for Ease of Use (80%) and Accuracy of Answers (65%). Relevance of Info was generally positive, while System Responsiveness and Transparency showed mixed results. Still, (60%) of users responded positively on the Recommendation dimension.
4. **FEMMa Oracle (17 participants):** The system performed well across all criteria. Accuracy was rated "accurate" or "highly accurate" by (64.7%), and Ease of Use by (82.3%). Relevance of Info was high (88.3%), and (88.9%) found it transparent. Responsiveness was rated "fast" by (50%) and "average" by (28.6%). Overall, (58.8%) said they would recommend it.
5. **Assist Doctor (21 participants):** Participants found the system easy to use, with (66.7%) rating it as "easy" or "very easy." Both Accuracy of Answers and Relevance of Info received favourable ratings, each at approximately (62%). System Responsiveness was positively reviewed by more than half of the users. About (62%) found the system transparent, and (47.6%) said they would recommend it.

Across the five systems, Ease of Use and Accuracy of Answers were consistently rated positively. Transparency and Recommendation showed more variation, sometimes due to design choices. For example, *Disarm RAG* used hidden source information. These differences show that user perception depends on the domain and output presentation.

## 6 Lessons Learned

While developing and evaluating the five RAG systems, we encountered technical, operational, and ethical challenges. These lessons reflect practical insights drawn from our hands-on engineering work, system deployment in real-world domains, and user-centred feedback. Together, they offer guidance for building reliable, domain-adapted RAG systems that balance performance, compliance, and trustworthiness.

### 6.1 Technical Development

Building RAG systems for real-world applications surfaced a number of technical hurdles that required hands-on problem solving and thoughtful design decisions.

- *Domain-Specific Models Are Essential*: General-purpose models like GPT-4o struggled with domain-specific and Finnish-language queries. Leveraging Finnish-optimized models like `Poro-34B`, along with compatible embedding models (e.g., `text-embedding-ada-002`), led to more contextually relevant responses.
- *OCR Errors Impact the Pipeline*: Noisy OCR output from agriculture and healthcare PDFs degraded FAISS quality. Using `TesseractOCR`, `easyOCR`, and regex-based cleanup improved extracted text.

- *Chunking Balances Speed and Accuracy*: Token chunk sizes between 200–500 struck a practical balance between retrieval relevance and query latency. Smaller chunks bloated the index, increasing lookup times.
- *FAISS Scalability Hits Limits*: With large corpora (>10k embeddings), FAISS latency increased noticeably. Metadata filtering by document type reduced search time.
- *Manual Environment Management*: Without containerisation, we faced version conflicts across PyTorch, FAISS, OCR libraries, and OpenAI APIs. Strict environment pinning and manual sync across development/production was necessary for stability.

### 6.2 Operational Factors

Operating RAG systems in real-world settings revealed practical challenges related to data workflows, infrastructure choices, and user interaction management.

- *SQLite for Tracking User Interaction*: We used SQLite to log user questions, responses, and ratings (e.g., in *AgriHubi*). This lightweight store helped identify system failures and understand user behaviour.
- *Scraping Pipelines Are Fragile*: Websites changed often, breaking parsers. Without stable APIs, we relied on semi-structured feeds and regular script maintenance.
- *Self-Hosted Setup for Speed and Compliance*: We hosted LLMs and vector stores on our own servers to reduce GDPR risks and improve speed. This approach balanced control with performance in sensitive domains.
- *Clean Data Boosts Retrieval Quality*: Removing OCR noise and duplicates from source data improved answer relevance without modifying models.
- *User Feedback Drives System Tuning*: User ratings and comments exposed weak spots, guiding adjustments to retrieval settings and chunk sizes.

### 6.3 Ethical Considerations

While technical and operational aspects were central to system performance, ethical considerations around transparency, and data bias proved equally important during deployment.

- *Source File References Build Trust*: Providing filenames and download links helped users validate AI outputs. In security use cases (e.g., *Disarm RAG*), sources were intentionally hidden to protect sensitive material.
- *Dataset Bias Impacts Retrieval Balance*: Unbalanced source data led to overrepresentation of some document types. Re-ranking improved diversity and fairness in answers.

**Practical and Research Takeaways:** Our findings highlight both persistent and emerging challenges in applying RAG systems to real-world, multilingual, and domain-specific settings. While issues like OCR noise, chunk size tuning, and retrieval balancing are well recognized, this study emphasizes the importance of practical strategies such as data cleaning, user feedback mechanisms, and lightweight response validation for improving retrieval quality and system reliability. These lessons extend current research by connecting it to deployment realities and offer value to the software engineering community by addressing concerns related to retrieval infrastructure, stability of data pipelines, and transparency in system outputs. These takeaways help guide the development of adaptable and trustworthy RAG solutions.

## 7 Study Limitations

This study presents findings grounded in the design, deployment, and evaluation of five domain-specific RAG systems, but several limitations must be acknowledged. First, while our evaluation involved 100 participants across diverse roles including researchers, practitioners, and domain experts, approximately 20% of the sample consisted of students. Although these students had relevant technical or domain experience, their feedback may reflect differing expectations or usage behaviour compared to full-time professionals. This demographic distribution, while broad, could influence the generalizability of findings to strictly industrial settings. Additionally, participants rated the accuracy of system-generated answers, yet only 17% identified as domain experts. Thus, accuracy ratings provided by non-experts might not reliably reflect the factual correctness of the systems' outputs.

Second, participants interacted with one or more systems, and survey responses were collected separately after each system use. Not all 100 participants engaged with every system; the number of responses per system varied based on individual interest and domain familiarity. For instance, feedback on *AgriHubi AI Assist* reflects only the users who selected and interacted with that system. This variation in exposure may affect the comparability of results across different systems, and the limited interaction time restricted analysis of longer-term user engagement.

Third, the lessons learned presented in this paper are based on our development experience and observations during system implementation and evaluation. While they do not result from formal empirical analysis, they reflect recurring challenges and design considerations encountered across multiple domains. Although not statistically validated, these insights can inform future work on the design and implementation of RAG systems in applied settings.

## 8 Conclusion

In this paper, we presented a tool-assisted approach for designing, implementing, and evaluating RAG-based systems across five real-world domains. Each system was tailored to its specific context–ranging from municipal governance to agriculture and healthcare by integrating multilingual OCR pipelines, semantic retrieval with vector embeddings, and either in-house or cloud-based LLMs. Our user study, involving 100 participants, provided insights into how these systems perform in practice, not just in terms of technical metrics, but also usability, transparency, and user trust.

Through our development work, we identified twelve lessons learned that highlight, in our view, recurring challenges in building practical RAG pipelines. These include balancing chunk size with latency, managing dependencies without containerization, and maintaining retrieval speed at scale. We also found that clean data, user feedback, and clear information presentation are critical for building trust. As industry and research interest in RAG systems grows [10,13], we hope these insights support future development efforts.

Looking ahead, we see a strong need for more structured evaluation mechanisms that go beyond user ratings. As future work, we propose integrating an *Evaluation Agent Model*, a system-internal module that checks AI-generated responses for accuracy, relevance, and completeness before presenting them to users. Based on our experiences, user feedback alone is often insufficient to catch factual errors or incomplete responses, especially in domains where missing or misleading information could have serious consequences. An automated evaluation agent could trigger second-stage retrievals or prompt reformulations when weaknesses are detected, creating an adaptive feedback loop. We believe that such mechanisms are essential to improving the reliability and trustworthiness of RAG systems in high-stakes, real-world applications.

**Acknowledgements.** This paper is based on research supported in part by the Synthetica project funded by the Research Council of Finland, the GENT project funded by the European Regional Development Fund, and the AgriHubi initiative supported by Pohjois-Pohjanmaan ELY-keskus (Centre for Economic Development, Transport and the Environment). The authors gratefully acknowledge this support. The authors declare no conflicts of interest related to this work.